# Helical and nonhelical (magneto-)Burgers turbulence: I. Compressibility reduction and beyond

Jian-Zhou Zhu (朱建州)[1*] and Pei-Xin Shi (史培新)[2]

[1] *Su-Cheng Centre for Fundamental and Interdisciplinary Sciences,*

*Gaochun, Nanjing 211316,*

*China and Fluid Institute,*

*Nanchang University,*

*Nanchang 330031, China*

[2]*Institute of Advanced Studies,*

*Nanchang University,*

*Nanchang 330031, China*

## Abstract

We compare the helical and nonhelical (magneto-)Burgers turbulence for the *helicity fastening effect*. Theoretical arguments and heuristic mathematical analysis are offered for the latter notion in the new system loosing some "nice" properties as previously used in addressing the Navier-Stokes and various plasma fluids. Miscellaneous discussions are also offered, including the inferences of several consequences on the transports of passive scalars for both the density and tracer, particularly, the opposite consequences of the helicity fastening effect for the latter two scalars in appropriate situations (with the caveat of the possibility of the inverse cascade of the tracer energy). Basic numerical results of the fractions of the parallel-mode spectra, with maximally-helical random forcing on some small-wavenumber modes, present a benefit of about 0.2 over those with nonhelical forcing, indicating regularization (to some degree) of the solutions. Such helicity "fastening" effect of Burgers turbulence is much more marked than that for low-Mach-number Navier-Stokes turbulence. The magnetic helicity in magneto-Burgers dynamics can present an even stronger benefit, of around 0.5+.

---

* jz@sccfis.org

# I. INTRODUCTION

The investigation of the effect of helicity on the statistics of the flow itself was initiated more than 60 years ago [1], while that on other fundamental problems is relatively new (e.g., Ref. [2] for the effect on the scalar and magnetic field transports). We are going to report relevant studies on both associated to the (magneto-)Burgers dynamics [3, 4].

Vorticity in the flows without the thermodynamic pressure, i.e., those governed by the Burgers equation, makes the latter "nonintegrable", in the sense that the Hopf-Cole transformation extended to higher dimension cannot be applied to linearize the system. Ohkitani and Dowker [5] compared the vortical Burgers and incompressible Navier-Stokes flows, but it is natural to expect closer relations, in many aspects, between the former and the compressible Navier-Stokes system. We double checked the earlier preliminary analysis of helical Burgers turbulence [6] and now report the results together with the coherent ones of magneto-Burgers turbulence, including other discussions (on the passive scalar transports, say). Ref. [5] sketched the arguments that suggest the Burgers equations be more singular than the Navier-Stokes equations, and we will show results indicating regularization (to some degree) of the solutions with helicity.

If we consider the degree of regularity, which can be defined by appropriate norms in some solution space, the fraction of compressive modes plays an important role. Thus, we may influence the regularity through the fraction of compressive modes. Such a possibility is supported by the consideration of the effects of helicity, i.e., those associated to the notion of "fastening turbulence" [7]. The latter refers to the result that, roughly speaking, helicities can reduce the fraction of compressive modes of neutral fluids and plasmas, which has been verified preliminarily in low-Mach-number fully developed turbulence [8]. So, helicity may regularize to some degree the flows (c.f., e.g., Refs. [9–12] for relevant rigorous analysis of incompressible flows, and Ref. [13] for heuristic remarks based on helical-mode interactions supporting the absolute equilibria). The helicity fastening effect indicates that the independent parameterizations of helicity and compressibility for the Kazantsev-Kraichnan model made in Ref. [2] then would not be complete, a good start though, if carried over to the Navier-Stokes or Burgers problems.

The helicity fastening effect for the Burgers dynamics appears unfavorable on the first sight, especially by simply checking the ideal conservation properties [14, 15] different to those used in the absolute equilibrium analysis [7]. But the frozen-in law of vorticity $\boldsymbol{\omega} = \nabla \times \boldsymbol{u}$, used in the geometrical argument with the chiral base flow/field (CBF) [7] (more on CBF in Sec. I B), still holds as in the (barotropic) Euler equation (see, e.g., the appendix of Ref. [5]). So, it deserves to perform the Burgers tests which shall be informative for the Navier-Stokes flows with large compressibility; similarly for the plasmas.

### A. Models

The Burgers equation reads

$$\partial_t \boldsymbol{u} + \boldsymbol{u} \cdot \nabla \boldsymbol{u} = \nu \nabla^2 \boldsymbol{u} + \boldsymbol{f}, \tag{1}$$

where the forcing $\boldsymbol{f}$ can be used to control the injection of helicity $\mathcal{H} = (2V)^{-1} \iiint_V \nabla \times \boldsymbol{u} \cdot \boldsymbol{u} \, dV$ in the domain of volume $V$: the flow is helical if $\mathcal{H} \neq 0$, otherwise nonhelical. With appropriate normalization, the kinetic viscosity $\nu$ would be simply the inverse of the Reynolds number. When the continuity equation for the mass density $\rho$ is included, different invariance laws involving the density $\rho$ and velocity $\boldsymbol{u}$ can be derived [14], but here we are only concerned with the properties of the self-autonomous velocity.

We also test the magnetic helicity effects on the reduction of compressibility in the solutions to the magneto-Burgers equation in which, for simplicity and to compare with the neutral case somehow on an equal foot, we take constant (unit) density in the Lorentz force (see also, e.g., Ref. [16] for such a treatment):

$$\partial_t \boldsymbol{u} + \boldsymbol{u} \cdot \nabla \boldsymbol{u} = \nu \nabla^2 \boldsymbol{u} + (\nabla \times \boldsymbol{b}) \times \boldsymbol{b} + \boldsymbol{f}_{kin}, \tag{2}$$

$$\partial_t \boldsymbol{b} = \nabla \times (\boldsymbol{u} \times \boldsymbol{b}) + \eta \nabla^2 \boldsymbol{b} + \boldsymbol{f}_{mag}. \tag{3}$$

Actually, we will compare the effects of magnetic helicity and kinetic helicity, injected respectively by the force $\boldsymbol{f}_{mag}$ and $\boldsymbol{f}_{kin}$. An interpretation of the vector $\boldsymbol{b}$ as the magnetic field is appropriate by imposing $\nabla \cdot \boldsymbol{b} \equiv 0$. For studies on magnetic field associated to helicity and chirality in models including magnetohydrodynamics (MHD), see recent reviews [17, 18] and references therein. Particularly, the model applied in a recent study on the existence of shear-current effects in magnetized Burgers turbulence [19] is different with only a Stokesian viscosity to the above magneto-Burgers system with the standard Burgers Laplacian viscosity.

## B. More physical motivations and plans

The terminology of "base flow" in "CBF" mentioned before is associated to the fact that a velocity vector field in the neighborhood of a location is generically (except for some possible singular points) of such a component-wise dimensional reduction that the velocity gradient matrix is of the "real Schur form" (more definite description in Sec. II A) in an appropriate co-ordinate system which can be obtained by an orthogonal transformation from a fixed frame [20]. Thus, a global flow/field with the gradient being uniformly (over some space-time domain) of the real Schur form ["real Schur flow"] may be used as a "base flow/field" as the cornerstone of an effective laminar flow diffeomorphism model, as well as of turbulence (but with additional elements). For example, to understand the above mentioned fastening effect of helicity, it is natural to think about the possibility of reduction to the helical real Schur flows (i.e., CBF), as proposed in Ref. [7] with particularly the notion of the coarse-grained (or "renormalized" by, say, arranging and collecting the appropriate Fourier modes) CBF structures. [Even for the turbulence in a cyclic box, the CBFs in a sub-domain or the whole domain does not necessarily be accordingly periodic (in all directions) for such a reductionism scenario [21]: the sub-domain situation is obvious, and for the whole domain one can think of the fact that a periodic function can be decomposed into non-periodic ones.]

Arnold and Khesin [22] remarked that the gauge groups exploited by physicists "occupy an intermediate position between the rotation group of a rigid body and the diffeomorphism groups. They are already infinite-dimensional but yet too simple to serve as a model for hydrodynamics." The "similarity" of the velocity gradients up to the transformation [$O(3)$ in 3-space] however indicates a possibility of a gauge theory and of reducing the diffeomorphism groups of hydrodynamics to some simplified model; and, particularly, the "real Schur frame" is reminiscent of the (local) inertial frame in general relativity theory where the Riemann metric can be transformed into a Minkowskii one. This can be more than an analogy but supported by the fluid/gravity duality associate to the notion of "holography" [23]. It follows then the possibility of developing a precise and deep correspondence between the local real Schur frame and the inertial frame (thus between the real Schur flow and special relativity), and, between the general (full-dimensional) Navier-Stokes flow and the *general relativity*, and, even further, the correspondence between the helicity "fastening" effect and the length contraction in *relativity* [6], among others. Systematically checking the relevance with holography and establishing a holographic (if indeed) or other principle unifying various turbulence physics needs a lot of insights and intuitions which can only be obtained from comprehensive investigations of the relevant flows, so in this series we carry the relevant ideas forward to the extreme cases of Burgers equation.

The consequences of helicity fastening effect are many, ranging from passive to active fields and particles (with and without inertial), and even for a particular category such as the passive transport, there can be various objects such as those corresponding to differential forms of different orders. The most "simple" (but highly nontrivial [24, 25]) are the passive scalar/tracer $c$ and density $d$ problems into which we may obtain some insights by basic analysis of the dynamics (Sec. II B). [Such transported matters are ideally "material" (i.e., Lie-transported geometrical objects/differential forms [22]: 0-form for $c$ and 3-form for $d$ multiplied by the volume form). They can have strong relevance to heavenly processes [26, 27], besides the well-known daily ones. Most recent developments of the transports of the passive vector corresponding to a 2-form, i.e., the linear dynamo problem, with field theoretic renormalization group technique and the operator product expansion may be found in Refs. [2, 28] and references therein (for also passive scalars), and for numerical studies of the effects on the transports of tracers by incompressible Navier-Stokes flows, see, e.g., Ref. [29].] Most complicated probably is the problem with strong back reaction onto the advecting flow, such as the nonlinear dynamo.

For the ideal magneto-Burgers system, the Lorentz force (per mass) presenting in the momentum equation prevents the vorticity from being material. With no frozen-in vorticity, we then in principle are not able to apply the associated Kelvin theorem essential to the geometry of the Taylor-Proudman effect to which the helicity effect was argued to be transformed [7]. So, again, on the first sight, the situation seems even worse for the kinetic helicity to take effect, for loosing further the geometrical argument. A similar problem in nonlinear dynamo has actually already been considered by Pouquet et al. [31] who noticed that the kinetic helicity can transform into the magnetic helicity. [Those authors discussed a specific situation with small-scale injection of kinetic helicity (and energy) associated to the inverse cascade of magnetic helicity, but the helicity transformation scenario can be reasonably generalized. That is, signed kinetic helicity (peaked around its injection scale) can lead to generation of the same-sign magnetic helicity at smaller scales and opposite-sign at larger scales. The inverse cascade is accompanied with the amplification of the opposite-sign magnetic helicity, thus a net amount of the latter.] So, there is at least one reason to expect the kinetic helicity work through the ideally invariant magnetic helicity to have a "fastening" effect as argued in Ref. [7] using the frozen-in magnetic field (associated to the Alfvén theorem).

In this first communication [32], we will numerically examine helicity fastening effect for both Burgers and magneto-Burgers dynamics in Sec. III, before which more theoretical considerations are laid out in Sec. II A below. Energy analyses for passive scalars are performed in Sec. II and conclusive remarks are offered in Sec. IV.

## II. THEORETICAL CONSIDERATIONS

### A. Physical arguments for the Burgers-flow helicity fastening effect

For simplicity, we consider flows in a cyclic box of dimension $2\pi$, with the Helmholtz-Hodge decomposition of a vector $\boldsymbol{v} = \boldsymbol{v}_\parallel + \boldsymbol{v}_\perp$ with $\nabla \times \boldsymbol{v}_\parallel = 0$ (and $\boldsymbol{u}_\parallel = \nabla \phi$ as will be used in Sec. II C) and $\nabla \cdot \boldsymbol{v}_\perp = 0$ for the "parallel" and "transversal" components, respectively, $\boldsymbol{v}_\parallel$ and $\boldsymbol{v}_\perp$ satisfying $|\boldsymbol{v}|^2 = |\boldsymbol{v}_\parallel|^2 + |\boldsymbol{v}_\perp|^2$. Eq. (1), with the viscosity and forcing set to zero for the time being, can be accordingly decomposed into those for the latter (see also Ref. [5]),

$$\partial_t \boldsymbol{u}_\parallel = -\nabla |\boldsymbol{u}|^2/2 + \boldsymbol{L}_\parallel, \tag{4}$$

$$\partial_t \boldsymbol{u}_\perp = \boldsymbol{L}_\perp, \tag{5}$$

with the Lamb vector $\boldsymbol{L} := \boldsymbol{u} \times \boldsymbol{\omega}$, which may most easily be seen in Fourier space (c.f., e.g., Ref. [30] for the Navier-Stokes equation). The local helicity density $h := \boldsymbol{\omega} \cdot \boldsymbol{u}/2$ is obviously competing with $\boldsymbol{L}$, in the sense of the complementary angles associated to the respective inner-product and cross-product between $\boldsymbol{\omega}$ and $\boldsymbol{u}$. So, we may expect that, in the higher helicity state with small $\boldsymbol{L}$-contribution, $-\nabla |\boldsymbol{u}|^2/2$ will reduce the $\boldsymbol{u}_\parallel$-energy. It turns out that favorable mechanism appears to be somehow analogous to that of the passive-scalar transports after the discussion of which in Sec. II B more analysis following Eqs. (4,5) will be offered in Sec. II C. We now turn to a physical scenario which is an extension of what we previously proposed [7] and which may be relevant to the effect of the $\boldsymbol{L}$-terms:

For Burgers flow with component-wise dimensional reduction such as $\partial_z \boldsymbol{u}_h \equiv 0$ for the horizontal velocity component $\boldsymbol{u}_h := \{u_x, u_y\}$ and the vertical component $u_z$ in general de-pending on all three coordinates $x$, $y$ and $z$ [thus a two-component-two-dimensional-coupled-with-one-component-three-dimensional (2D2Dcw1C3D or simply 2D2D3D) real Schur flow], a sub-system of $\boldsymbol{u}_h$ is governed by a self-autonomous two-dimensional Burgers equation $(\partial_t - \nu\nabla^2 + \boldsymbol{u}_h \cdot \nabla)\boldsymbol{u}_h = 0$ [33]. The "coupling" for $u_z$ is "passive", without feedback onto $\boldsymbol{u}_h$; that is, $u_z$ becomes a nonlinear passive scalar. The helicity contributed by $\nabla \times \boldsymbol{u}_h \cdot \boldsymbol{u}_v$, with the vertical component $\boldsymbol{u}_v := \hat{\boldsymbol{z}} u_z$ along the (unit) $\hat{\boldsymbol{z}}$ direction, then should not have "authentic" effect on $\boldsymbol{u}_h$ (except that the evolution of the horizontal component depends on the initial field and the forcing that controls helicity), however, we note that now the three-dimensional (3D) spatial variation of $\boldsymbol{u}_v$ also contributes to the helicity and can dynamically reduce the $\partial_z u_z$ contribution to the parallel mode. Also, for a 3D isotropic turbulence, (magneto-)Burgers or Navier-Stokes, the reduction to real Schur flows may actually still have the other possibility: we have still the two-component-three-dimensional-coupled-with-one-component-one-dimensional (2D3Dcw1C1D or simply 3D3D1D) real Schur flow of the type with $\partial_x u_z \equiv 0 \equiv \partial_y u_z$ corresponding to one of the real-Schur-form duo of a (velocity gradient) matrix [34], and such a helical 3D3D1D real Schur flow, which may also be called a CBF, can also present helicity fastening effect from the three-dimensional $\boldsymbol{u}_h$.

There can be two different points of view on the above issue: one is that since we can always locally transform the velocity gradient into one chosen real Schur form, then a particular type of real Schur flows, say, the 2D2D3D one (as chosen in Ref. [7]), should be sufficient; the other is that there should be no reason to exclude either one. Actually, numerical results as partly reported in [35] show that the horizontal two-dimensional (2D) solenoidal-mode inverse energy transfer mechanism dominates to support the large-scale vortexes, with the other modes presenting forward energy transfers. Navier-Stokes full-dimensional flow in 3-space without boundary and other particular constraints however is dominated by forward energy transfers. Thus, we may indeed need to consider the (helical) 3D3D1D real Schur flow as the (C)BF alternative to the 2D2D3D one, to complete the bases supporting the reductionism of turbulence, and the uncertainty between the duo can actually facilitate the introduction of randomness into the model for turbulence.

In the frame with "mean rotation" rate $\Omega$ in the direction $\boldsymbol{R}$, the new velocity $\boldsymbol{u}' = \boldsymbol{u} - \boldsymbol{\Omega} \times \boldsymbol{r}$ and vorticity $\boldsymbol{\omega}' = \boldsymbol{\omega} - 2\boldsymbol{\Omega}$, and thus $\boldsymbol{\omega} \cdot \boldsymbol{u} - \boldsymbol{\omega}' \cdot \boldsymbol{u}' = \boldsymbol{\Omega} \cdot (2\boldsymbol{u} + \boldsymbol{r} \times \boldsymbol{\omega})$ the integration of which over space on both sides resulting in vanishing helicity in this frame leads to $\mathcal{H} = \boldsymbol{\Omega} \cdot \int_V [2\boldsymbol{u} + \boldsymbol{r} \times (\nabla \times \boldsymbol{u})] d^3\boldsymbol{r}/2V$ for determining $\Omega$. So, the same geometrical essence of Taylor-Proudman effect about reducing the compressibility in the rotating plane works in principle for more general $\boldsymbol{R}$, not necessarily the vertical direction of the 2D2D3D real Schur flow, and for more general flows; and, with $\boldsymbol{R}$ not determined *a priori*, the effect can be contributed from different directions. Choosing $\boldsymbol{R}$ to be vertical has the merit that it is connected with a kind of gauge invariance of the real Schur form: both 2D2D3D and 3D3D1D real Schur flows are respectively preserved to be of the same type by any transformations such as diffeomorphisms [e.g., most simply, $U(1)$] in the horizontal plane, which might indicate a deep justification for vertical $\boldsymbol{R}$. So, the reduction of the helicity fastening effect on isotropic turbulence to that on 2D2D3D CBFs in [7] is modified to include also the 3D3D1D ones.

The particular studies of 2D2D3D and 3D3D1D Burgers flows have been initiated for a specific communication which will be a follow-up in this series [36].

### B.  Other consequences: transports of (passive) tracer and density scalars

A passive tracer which may be the (perturbations of) temperature $c$ satisfies

$$(\partial_t - \chi_c\nabla^2)c - \varphi_c = -\boldsymbol{u}\cdot\nabla c = -\nabla\cdot(c\boldsymbol{u}) + c\nabla\cdot\boldsymbol{u}, \qquad (6)$$

with $\varphi_c$ the corresponding pumping and $\chi_c$ the diffusive coefficients. Passive scalar energy $\langle c^2 \rangle$ has been argued to present inverse cascade to large scales (approaching Gaussian statistics with no intermittency) in the Kraichnan model where some precise results of multi-point correlations can be computed, with the degree of compressibility above a critical value [24]. For Burgers (not Kraichnan's Brownian velocity field white in time) flow, so far, we neither know the critical value nor precisely understand how the reduction of the absolute value of helicity increases the compressibility to that value for an inverse cascade [37]. But, it means that the "fastening" effect on the degree of compressibility affects the scalar energy transfers (favoring more the forward ones), if not abruptly change the direction of cascade.

We see that the velocity divergence or correspondingly the parallel/compressive mode also presents in the term in addition to the local divergence of the flux in the right-hand side of Eq. (6). It is this term which makes it different to the transport of a density $d$ (c.f., e.g., Ref. [24]) which, ideally, is of the local conservative form,

$$(\partial_t - \chi_d\nabla^2)d - \varphi_d = -\boldsymbol{u}\cdot\nabla d - d\nabla\cdot\boldsymbol{u} = -\nabla\cdot(\boldsymbol{u}d), \qquad (7)$$

with $\varphi_d$ being the pumping and $\chi_d$ the corresponding diffusion coefficients, or it is the second term in the middle of Eq. (7) that makes the latter different to that of *c*: the helicity "fastening" effect then can reduce these terms (and thus the corresponding differences between the *c* and *d* dynamics) in the respective case, which is the **Consequence** No. 1 of the helicity fastening effect.

We can further derive other results. For example, we have the scalar energy transport equation

$$(\partial_t - \chi_c \nabla^2)\frac{c^2}{2} + \chi_c(\nabla c)^2 - \varphi_c c = -\nabla \cdot \frac{\boldsymbol{u}c^2}{2} + \frac{c^2}{2}\nabla \cdot \boldsymbol{u}, \qquad (8)$$

$$(\partial_t - \chi_d \nabla^2)\frac{d^2}{2} + \chi_d(\nabla d)^2 - \varphi_d d = -\nabla \cdot \frac{\boldsymbol{u}d^2}{2} - \frac{d^2}{2}\nabla \cdot \boldsymbol{u}. \qquad (9)$$

Opposite signs present in the last terms of the right-hand sides of Eqs. (8) and (9), and the different correlations between the transported matters and the velocity divergence can result in different effects. Without these last terms, those equations are ideally of the local conservation form. Integrating over the space through whose boundary no net fluxes pass, we see that these last terms involving velocity divergence directly take effect. To be more explicit, we can write down the equation with spatial and ensemble averages (denoted respectively by the overline, "$\overline{\bullet}$", and angle brackets, "$\langle\bullet\rangle$"):

$$(\partial_t - \chi_c \nabla^2)\frac{\langle\overline{c^2}\rangle}{2} + \chi_c\langle\overline{(\nabla c)^2}\rangle - \langle\overline{\varphi_c c}\rangle = \frac{\langle\overline{c^2 \nabla \cdot \boldsymbol{u}}\rangle}{2}, \qquad (10)$$

$$(\partial_t - \chi_d \nabla^2)\frac{\langle\overline{d^2}\rangle}{2} + \chi_d\langle\overline{(\nabla d)^2}\rangle - \langle\overline{\varphi_d d}\rangle = -\frac{\langle\overline{d^2 \nabla \cdot \boldsymbol{u}}\rangle}{2}. \qquad (11)$$

With the shocks skewing the probability distribution function of the velocity derivative on the negative side [3], the large negative velocity divergence comes with the concentration of *d*. We then can expect that, "overall" or "statistically speaking", $-d^2\nabla \cdot \boldsymbol{u}$ is more positive than $c^2\nabla \cdot \boldsymbol{u}$, or at least less negative, resulting in higher level of $\langle\overline{d^2}\rangle$ than $\langle\overline{c^2}\rangle$ when the pumping and diffusivity effects are the same: **Consequence** No. 2 of the helicity fastening effect.

$\nabla \cdot \boldsymbol{u}$ is overall less negative with larger helicity, i.e., the last term $\frac{c^2}{2}\nabla \cdot \boldsymbol{u}$ is overall more positive, resulting in relatively higher level of $\langle\overline{c^2}\rangle$ than the less or non-helical case; and, accordingly for the case of the transported density, with lower level of $\langle\overline{d^2}\rangle$ for more helicity (opposite to the *c*-case): **Consequence** No. 3 of the helicity fastening effect.

The above reasoning however is (quasi-)linear, assuming that the dynamics in comparison are "close", which may not be the case due to other effects that can lead to the abruptly different possibility of the inverse cascade of $\langle \overline{c^2} \rangle$ for $c$ transported by compressive enough velocity [39]. But, if the helicity injection (on sufficiently many modes) is large enough to have insufficient compressibility for inverse cascade of the tracer energy, or, if the pumping of the tracer is already at the largest scales, such an inference becomes more favorable and the results can be used to compare with the cases with helical but less-helical forcing.

When the $c$-energy inverse cascades happen in two cases different mainly in the amounts of helicity, further normalization by, say, the appropriate norm of the first term on the right hand side of Eq. (8) to reasonably screen out the enhancement of inverse cascade (thus, presumably the energy level of $c$) with less helicity, the $c$-energy should still have a lower level, as **Consequence** No. 3 of the helicity fastening effect in the above claims; otherwise, if the effect of the second term on the right hand side of Eq. (8) is screened out by some appropriate normalization, the opposite consequence results, and we call this **Consequence** No. 4 of the helicity fastening effect.

The above four inferences of the **Consequences** of the helicity fastening effects on the tracer and density transportations are basic and heuristic, while more involved calculations of the simultaneous (and independent) influences of helicity and compressibility on anomalous scaling of the tracer in the Kraichnan model (as well as the magnetic field in the Kazantsev-Kraichnan model) have been performed in Ref. [2]. Numerical studies are expected to be helpful to test the relevant assumptions, approximations and results [38]. There are also many other problems, such as particle acceleration (plasma heating or cosmic ray acceleration) and vector transport (the magnetic field in the linear dynamo), which cannot be exhausted here. And, as mentioned in the context of magneto-Burgers-dynamics, the comparison to active fields (e.g., Ref. [25]) for the issue of fastening effect would also be illuminating.

### C. (Magneto-)Burgers energy dynamics for the helicity fastening effect

We now turn to search for the analytical insights of the helicity fastening effect for Burgers turbulence itself. From Eqs. (4, 5), hiding again the forcing and viscous damping terms, we immediately obtain

$$\partial_t \langle \overline{|\boldsymbol{u}_\parallel|^2} \rangle = \langle \overline{(|\boldsymbol{u}_\parallel|^2 + |\boldsymbol{u}_\perp|^2)\nabla \cdot \boldsymbol{u}_\parallel} \rangle + \langle \overline{2\boldsymbol{u}_\parallel \cdot \boldsymbol{L}_\parallel} \rangle, \tag{12}$$

$$\partial_t \langle \overline{|\boldsymbol{u}_\perp|^2} \rangle = \langle \overline{2\boldsymbol{u}_\perp \cdot \boldsymbol{L}_\perp} \rangle, \tag{13}$$

corresponding to respectively the parallel and transversal energy equations: see Sec. II B for similar formulation of the scalar transport problem. [The ensemble average, denoted by $\langle \bullet \rangle$, is not explicitly necessary in the following reasoning, but since the discussion is about the dynamics, i.e., the time evolution, the resulted effect to be pointed out is not supposed to be apparent in the instantaneous realization at each time, but accumulated over time or statistical samples, thus the necessity of $\langle \bullet \rangle$.] The first term on the right hand side of Eq. (12) should be, thus an "intrinsic sink", due to the fact that the shocks skew the probability distribution function (PDF) with large (negative) velocity divergence [3] ($\nabla \cdot \boldsymbol{u}_\| = \nabla \cdot \boldsymbol{u}$) and $u_\|^2$.

The local helicity density $h := \boldsymbol{\omega} \cdot \boldsymbol{u}/2$ being about inner-product and the Lamb vector cross-product, we may start considering the ideally extreme case for maximal helicity with $\boldsymbol{\omega}$ being parallel to $\boldsymbol{u}$ almost everywhere (c.f., Sec. IV for remarks on anti-parallelity), thus vanishing $\boldsymbol{L}$. Compared to the transversal energy, this additional term in Eq. (12), $\langle \overline{(|\boldsymbol{u}_\||^2 + |\boldsymbol{u}_\perp|^2)\nabla \cdot \boldsymbol{u}_\|} \rangle$, dominates to maximally reduce the parallel energy; and, for isotropic turbulence solenoidally forced at large scales, the otherwise non-vanishing $\boldsymbol{L}_\|$ term is supposed to be positive and generate the large-scale parallel energy, while the $\boldsymbol{L}_\perp$ term should be negative to transfer the injected transversal energy out: most of the energies are contained at large scales for both the parallel and transversal components, so effectively the fraction of the parallel energy is expected to increase, a consistent argument in favor of the smaller fraction of the parallel energy in the helical state.

[Corresponding to the Navier-Stokes turbulence, there will be another pressure gradient term on the right-hand side of the parallel-energy Eq. (12), i.e.,

$$-\langle \overline{2\boldsymbol{u}_\| \cdot \nabla \ln \rho} \rangle = -\langle \overline{2\nabla\phi \cdot \nabla \ln \rho} \rangle \qquad (14)$$

where, for simplicity and easy analytical trackability, the isothermal approximation (with the pressure $p = c^2\rho$), $\boldsymbol{u}_\| = \nabla\phi$ and appropriate normalization leading to unit sound speed $c$ and background density $\rho_0$ have been made use of (c.f., e.g., Ref. [7]). Applying Green's first identity,

$$\iiint_V \nabla\phi \cdot \nabla \ln \rho \, dV = \iint_{\partial V} \ln \rho \nabla\phi \cdot d\boldsymbol{S} - \iiint_V \ln \rho \nabla^2 \phi \, dV,$$

we have

$$V\overline{\boldsymbol{u}_\| \cdot \nabla \ln \rho} = \iint_{\partial V} \ln \rho \nabla\phi \cdot d\boldsymbol{S} - V\overline{\ln \rho \nabla \cdot \boldsymbol{u}_\|} = \iint_{\partial V} \ln \rho \boldsymbol{u}_\| \cdot d\boldsymbol{S} - V\overline{\ln \rho \nabla \cdot \boldsymbol{u}} : \qquad (15)$$

the first surface integration term of the latter is a parallel flux across the domain and should vanish (indicated by a slash) for flows in a cyclic box or with fixed solid boundary etc., and the second term, according to the dilatation relation between the density and velocity divergence, is positive on average. Thus, the averaged pressure-gradient work rate on the parallel energy in Eq. (14),

$$-2\langle\overline{\boldsymbol{u}_\parallel\cdot\nabla\ln\rho}\rangle = 2\langle\overline{\ln\rho\nabla\cdot\boldsymbol{u}}\rangle, \quad (16)$$

is negative. With such additional enhancement of negativity, the importance of the **L**-term effect associated to helicity discussed in the above is relatively weakened; that is, the helicity fastening effect is understood to be weaker than that in the Burgers dynamics.]

We have not been able to have more quantitative computation for the stronger effect with larger helicity but believe that the above analysis, so far somewhat hand-waving though, be relevant for a rigorous proof, and a detailed comparison with the scalar energy dynamics will be illuminating [38]. Note that for magneto-Burgers turbulence, the Lorentz force $\nabla\times\boldsymbol{b}\times\boldsymbol{b}$ should be included in **L**. Magnetic helicity belongs more to the large scales, realizable by inverse cascade (when injected at small scales [17]) or subjected to less small-scale damping with weaker forward transfer, which is understood to increase the effect of raising $\langle\overline{|\boldsymbol{u}_\parallel|^2}\rangle$ and reducing $\langle\overline{|\boldsymbol{u}_\perp|^2}\rangle$ by turning on and strengthening the **L**-terms mentioned in the above, thus in turn appearing to favor the more effective fastening action.

## III. NUMERICAL RESULTS

### A. Burgers turbulence

#### 1. Simulation setups

Preliminary numerical tests of the full-dimensional and component-wise dimensionally reduced Burgers equations had been performed [6] with everything else the same as the Navier-Stokes and real Schur flows computed in Ref. [35]. Aside from some minor quantitative differences which are not relevant to our conclusions, those Burgers results from computations with different methods and discretization schemes agree with later ones from the computations with the modifications[40] of the PENCIL CODE [41]. The latter solver is more readily modifiable for the magneto-Burgers simulations to be analyzed below, so we describe the main elements of it for the results to be presented:

Sixth-order central-difference scheme is used for spatial discretization and third-order Runge-Kutta for time marching. The nonhelical forcing of the velocity is of the Taylor-Green type (parameterized by a phase $\theta$ [42]), written in the $x$-$y$-$z$ coordinate frame,

$$f_x^{\text{nh}} = \frac{A_{\text{kin}}}{\sqrt{3}} \sin\left(\theta + \frac{2\pi}{3}\right) \sin x \cos y \cos z, \tag{17}$$

$$f_y^{\text{nh}} = \frac{A_{\text{kin}}}{\sqrt{3}} \sin\left(\theta - \frac{2\pi}{3}\right) \cos x \sin y \cos z \tag{18}$$

$$f_z^{\text{nh}} = \frac{A_{\text{kin}}}{\sqrt{3}} \sin\theta \cos x \cos y \sin z, \tag{19}$$

with $\theta$ randomly chosen uniformly over $[0, 2\pi)$ at each time step and $A_{kin} = 60$ in our simulations. The purely helical sector [43] $\boldsymbol{f}^{\text{hel}}$ of such an $\boldsymbol{f}^{\text{nh}}$ reads,

$$\boldsymbol{f}^{\text{hel}} = \frac{1}{\sqrt{2}} \boldsymbol{f}^{\text{nh}} + \frac{1}{\sqrt{6}} \boldsymbol{\nabla} \times \boldsymbol{f}^{\text{nh}}. \tag{20}$$

$\nu = 7.5e-3$ and the final Reynolds numbers based on the root mean squares of the velocities are about several tens (depending on the choices of the parameters for definition), with differences smaller than 5% which is negligible and can be treated as identical for our analysis below in terms of spectral ratio, a kind of further self-normalization. [Further or alternative normalizations may be even "fairer" in some sense for the comparison, but, given the close Reynolds numbers and dissipative wavenumbers (in the sense of Kolmogorov), they are not necessary for our purpose here to painlessly modify the marked fastening effect (see below).] All the computations are in a box of $2\pi$ periods, with $N^3$ grid points homogeneously arranged.

Note that we have checked results from various Reynolds numbers and resolution $N$, presenting consistent results and showing the current results with Reynolds numbers and $N = 168$ are sufficient for our purpose here.

In a statistically steady state over which we collect data and take time averaging (also denoted by $\langle \bullet \rangle$ with the assumption of ergodicity), we define the power spectra as

$$E_{\text{kin}}(k,t) = \frac{1}{2} \sum_{|\boldsymbol{k}|=k} |\hat{u}(\boldsymbol{k})|^2, \tag{21}$$

where variables wearing hats indicate their Fourier transforms. $\sum_k E_{\text{kin}} = \overline{u^2}/2 := \mathcal{E}_{\text{kin}}(t)$, with the overbar $\overline{\bullet}$ denoting a volume average as used before. Additionally, we define the power spectrum of the compressive mode of $\boldsymbol{u}$ as

$$E_\parallel(k) = \frac{1}{2} \sum_{|\boldsymbol{k}|=k} |\hat{\boldsymbol{u}}(\boldsymbol{k}) \cdot \boldsymbol{k}/k|^2. \qquad (22)$$

And we have $\sum_k E_\parallel = \overline{u_\parallel^2}/2 := \mathcal{E}_\parallel(t)$.

*2. Results*

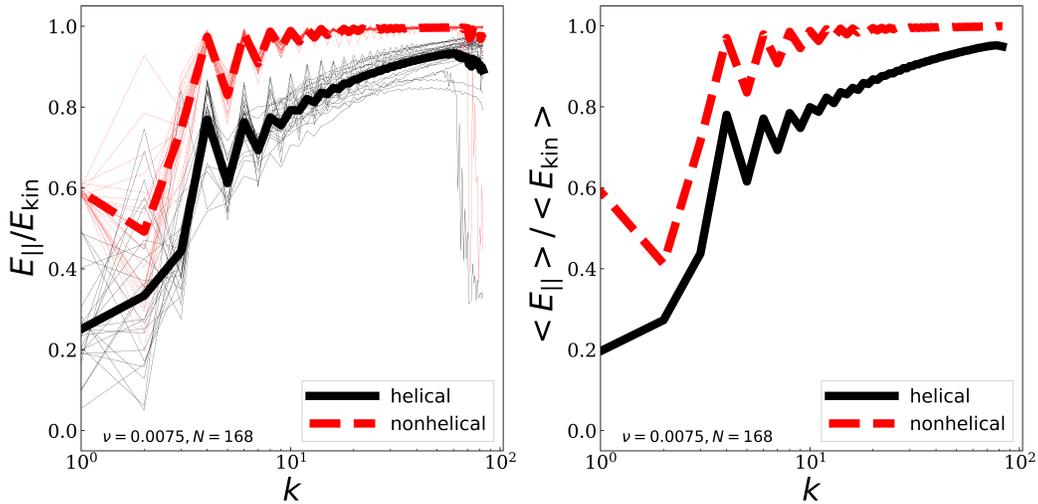

FIG. 1. Fractions of parallel modes: the average of the spectral ratios (left panel: solid thick line for the nonhelical case and dashed thick line for the helical case, with thinner lines respectively of the same type for 20 of a large number of snapshots to demonstrate the corresponding fluctuations) and the ratio of averaged spectra (right panel: also solid line for the nonhelical case and dashed line for the helical case), both showing an overall benefit of more than about 0.2+ in the helical case, with minor differences.

Fig. 1 compares the fractions of the compressive-mode energy over the total kinetic energy at wave number $k$, defined as $r^\parallel = E_\parallel/E_{\rm kin}$. The overall deficit, between $_h$elical and $_n$on$_h$elical cases, $\langle r_{\rm h}^\parallel - r_{\rm nh}^\parallel \rangle \approx -0.2$ over a wide range (except the near end of the dissipation range) confirms our expectation that helicity reduces the turbulent compressibility of the flow. [The benefits with such forcing schemes also mildly depend on the Reynolds numbers, and the value 0.2 can be raised by increasing the helicity injection (on more modes, say, at small-$k$ regime).] The meaningful comparison is the average but the fluctuations over time of *r*s are also presented in the left panel to show that the benefits are systematic, for almost every sample (except for some at the forced wavenumber $k = \sqrt{3}$).

A different definition of the ratio, i.e., the ratio of averaged spectra is also used in the right panel, showing negligible differences for our purpose.

Further appropriate normalizations (such as that by the Reynolds numbers, which are close here, as mentioned earlier) could be introduced and should be a small modification here, without changing the conclusion of the fastening effect. The spectral ratio in the dissipation range is somewhat subtle (sensitive to numerical accuracy etc.) but essentially irrelevant for our purpose [44]: the dissipation range contains a very small part of the total energy, and a global measure $\langle \mathcal{E}_\| / \mathcal{E}_{\text{kin}} \rangle$ or $\langle \mathcal{E}_\| \rangle / \langle \mathcal{E}_{\text{kin}} \rangle$ is found to present a benefit of 0.2+; the latter is much more marked/stronger than that in Navier-Stokes turbulence seen in Ref. [8], which seems surprising on the first sight but actually can be reasonably understood as remarked in Sec. II C.

### B. Magneto-Burgers flows

#### 1. Numerical details

Magneto-Burgers flows governed by Eqs. (2, 3) are simulated similar to those of Burgers in Sec. III A. The magnetic field is computed through the potential vector $\bm{a} = (\nabla\times)^{-1}\bm{b}$ satisfying $\partial_t \bm{a} = \bm{u} \times \bm{b} - \eta \bm{j}$ with $\bm{j} = \nabla \times (\nabla \times \bm{a})$. For all the runs (Table I), we use unit Prandtl number with and $\nu = \eta = 7.5 \times 10^{-3}$. The helical and nonhelical forcing for both $\bm{f}_{kin}$ and $\bm{f}_{mag}$ are implemented according to Eqs. (17–20) respectively, independently.

TABLE I. Summary of the magneto-Burgers runs.

| Run | $f_{\text{kin}}$ | $A_{\text{kin}}$ | $f_{\text{mag}}$ | $A_{\text{mag}}$ |
|---|---|---|---|---|
| KnhMnh | nonhelical | 60 | nonhelical | 60 |
| KhMnh | helical | 60 | nonhelical | 60 |
| KnhMh | nonhelical | 60 | helical | 60 |
| KhMh | helical | 60 | helical | 60 |

*2. Results*

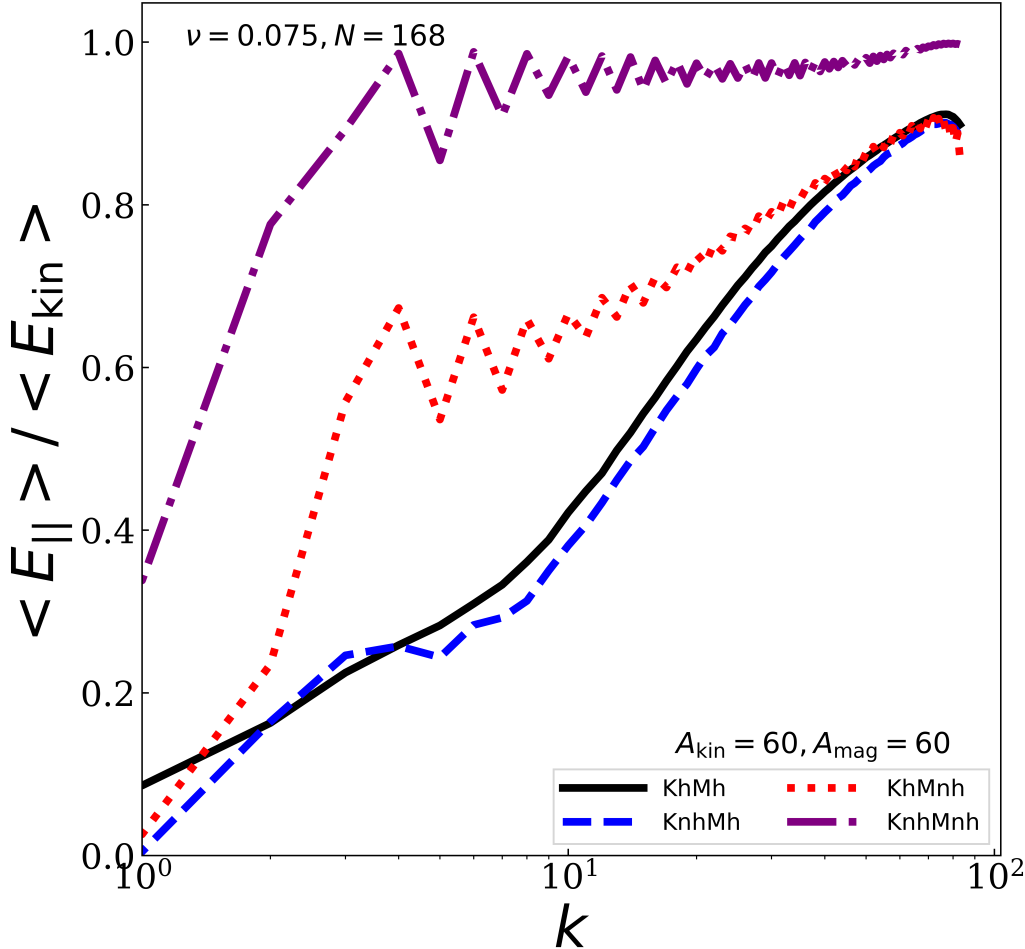

FIG. 2. The spectral ratios $\langle E_{\parallel}\rangle/\langle E_{kin}\rangle$ and $E_{kin}$ of magneto-Burgers runs.

Fig. 2 presents the averaged spectral ratios. Like the neutral Burgers case, an obvious difference on the $k$ dependence of the parallel-mode spectral ratios, between the Burgers results here and those of Navier-Stokes in Ref. [8], is that the ratio monotonically grows, in the (nearly-)inertial range and the beginning part of the dissipation range, for the former case but decays in the latter. Since the asymptotic inertial-range scaling law for the compressive-mode energy spectrum is $k^{-2}$ when the shock structures dominate, that for the vortical mode then should be even steeper, which might be regarded as a kind of ("external") intermittency correction to the $k^{-5/3}$ scaling: the "internal" intermittency, if exists, in the inertial range of incompressible Navier-Stokes turbulence should be much smaller than $2-5/3 = 1/3$ and not enough to make the exponent steeper than $-2$, while the "body force" at all scales from the interactions with the compressive modes of (highly) compressible turbulence may sufficiently change the (dimension of) the dissipative structures of the solenoidal component. To the best of our knowledge, there are no experimental or numerical evidences that exclude the possibility of a scaling exponent steeper than $k^{-2}$ for the solenoidal component of a (highly compressible) turbulence. The log-log plot (not shown) of the left panel, as also for that of Fig. 1, seems to indicate different scaling laws in the inertial range of the helical and nonhelical cases.

The magnetic helicity presents a spectral-ratio benefit of more than about 0.5 in the main body of the spectra, an apparently stronger "fastening effect" than that of kinetic helicity in Burgers turbulence (Sec. III). [Again, the benefits with such forcing schemes also mildly depend on the Reynolds numbers, and the value 0.5 can be raised by increasing the (magnetic) helicity injection (on more modes, say, at small-$k$ regime). It is possible to make a more precise comparison between the fastening effects of kinetic helicity and magnetic helicity by quantify the values of helicities, but, with these two helicities being in a sense coupled (see below) and for lack of systematic theory, we are content with such an intuitive rough one with the understanding given in Sec. II C.] When magnetic helicity is working, the kinetic helicity effect does not present clearly, obviously no linear superposition. Such a "saturation" of the effect from magnetic helicity is not surprising according to the analysis of MHD in Ref. [7] (where magnetic helicity but not the kinetic helicity is ideally invariant). Also, as remarked in Sec. I B, the kinetic helicity tends to generate opposite-sign (net) magnetic helicity (see, also, e.g., Ref. [45] on the picture for different Prandtl numbers, and references therein for other relevant details): the classical scenario is associated to the so-called $\alpha$-dynamo with kinetic helicity injected at large $k_f$, which, as checked in our data, however can be well generalized, even to our case with small $k_f$ ($=\sqrt{3}$) and with also magnetic forcing, simply according to the fundamental mode interactions. Such details are not the focus of this work and will be addressed in the future to compare with the passive-vector transport, i.e., the kinetic/linear dynamo problem (preliminary results with large $A_{kin}$ : $A_{mag}$ indicate some different relative importance of kinetic and magnetic helicity). In our simulations for "KhMh" and "KnhMh" in Fig. 3 (left panel), since kinetic and magnetic helicity injections at $k_f = \sqrt{3}$ are of the same (statistical) property, kinetic helicity injection tends to cancel a bit of magnetic helicity and thus the effect of the latter. The differences are small: even more subtle details associated to the secondary "fastening" effect can be involved with the **b** affected by the "fastened" **u** which is then in turn affected by the Lorentz force.

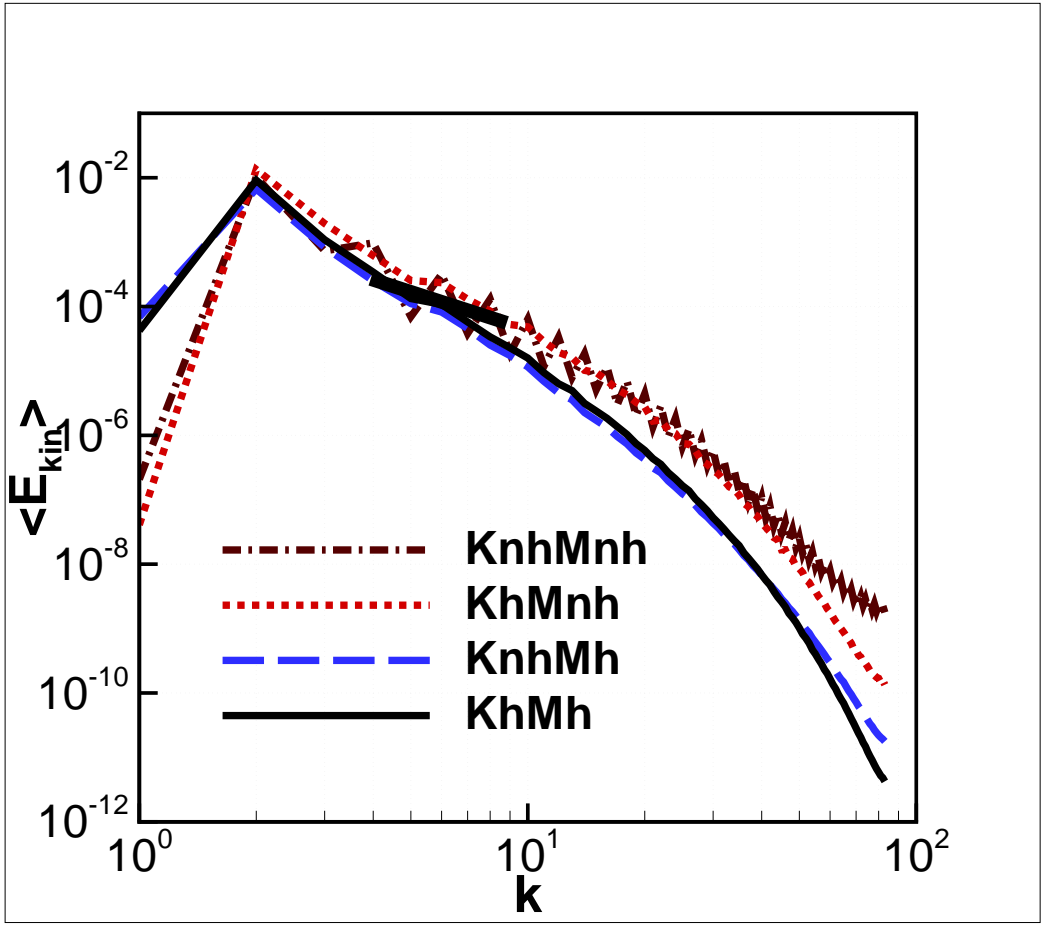

FIG. 3. The average spectra $\langle E_{kin} \rangle$ of magneto-Burgers runs, with the $k^{-2}$ law (thicker short line) indicating shocks added for reference.

Fig. 3 presents the averaged kinetic energy spectra. Consistent with the helicity fastening effect, the spectra with nonhelical forcing are closer to a shock-signaling $k^{-2}$-scaling in the (nearly-)inertial ranges: nonhelical cases allow more compressive modes which tend to form more/stronger shocks.

It would also be interesting to check the corresponding situations with magnetic helicity injected at small scales and with different combinations of the signs of helicity injections. Cross-helicity also can have the fastening effect in MHD [7], among others, which may be carried over to magneto-Burgers-dynamics and tested, say, by using appropriate $\boldsymbol{f}_{kin}$ and/or $\boldsymbol{f}_{mag}$ schemes. Such a comprehensive numerical studies in different setups however are beyond the scope of this note.

## IV. CONCLUSION, REFLECTION AND EXPECTATION

Our key conclusion, supported by both new theoretical arguments, heuristic analysis and numerical evidences, is that the helicities in (magneto-)Burgers flows present "fastening" effect, i.e., reducing the compressive modes, with benefits much more marked than in the low-Mach-number Navier-Stokes turbulence; and, the magnetic helicity in magneto-Burgers flows appears to have even much stronger such effect. A completely rigorous quantitative calculation of the helicity fastening effect is not available, but we hope the work presented here and later with various new facets can be motivating and illuminating for eventually even more systematic or ultimately thorough analyses, or, for other analytical approximations (e.g., the perturbative ones [28]).

Note that the parameterizations of helicity and compressibility in the two-loop approximation with the field theoretic renormalization group and operator-product expansion in Ref. [2] are independent, while we are emphasizing the effect of the former on the latter, i.e., the nonlinear dynamics of the advecting velocity, and the subsequent results on transportations. The perturbative approaches have been carried over to the problems of transports by the Navier-Stokes flows [28] (and references therein), which would be interesting to be further carried over to Burgers flows including the information of helicity and compressibility and compared with (nonperturbative) numerical results [36] (see, e.g., Ref. [47] for such an effort). Also, it is possible to construct a useful model, taking the helicity fastening effect (relating the spatial parity violation and compressibility) into account, beyond the celebrated (Kazantsev-)Kraichnan one to carry, say, the analytical approach of Ref. [2] further to be physically even more realistic.

We have only reported results of driven turbulence for the statistically steady states, but the fastening-effect issue in decaying compressible Navier-Stokes turbulence [46] and decaying (magneto-)Burgers turbulence, with the effect exposed in some different way, say, the late-time decaying laws, is also interesting.

This work is also in a sense programmatic, with brief analyses and results on magneto-Burgers flows to demonstrate the possibility of various other extensions (e.g., to different plasma fluid models including the extended MHD and two-fluid models without thermal pressure). Four conjectures on passive-scalar transports implies various other consequences in different physical situations (e.g., the magnetic field in the linear dynamo problem).

Part of our previous arguments [7, 8] exploit invariance laws which, *a priori*, no longer formally hold for the ideal (magneto-)Burgers dynamics. It is not impossible that they hold to a sufficient degree during some important and relevant dynamical process, but it may also be that it is some other aspects, such as the chirality, in the presence of helicity that really matter, which was partly the reason for the speculation on helical Burgers in Ref. [30]. One aspect of the presence of helicity is the loss of parity invariance, i.e., the invariance of the dynamics with respect to the reflexion of spatial coordinates. The latter was found to cause anisotropic large-scale instability even when the helicity is vanishing [48]. Note that $L$ vanishes when $\omega$ and $u$ are parallel or anti-parallel, the respective $h$s are of opposite signs, cancelling each other: such nonhelical state are very special and not supposed to be generic in nonhelical turbulence, thus not seriously challenging the analytical heuristic argument in Sec. II C for the Burgers helicity fastening effect. Another possible aspect is that, topologically speaking, helicity means nontrivial knottedness which however can present without helicity [22]: an appropriately designed comparison remains to be performed, say, with some techniques of vortex tube construction [49]. Thus, much more remains to be investigated for relevant issues of (magneto-)Burgers turbulence, including the deep physics of higher-order statistics (e.g., Refs. [2, 50, 51]).

As remarked in Sec. II A, a component-wise dimensionally reduced flow of the Burgers equation somehow pushes the turbulence reductionism idea to an extreme situation, calling for more critical tests. We expect systematic studies on both Burgers and Navier-Stokes component-wise dimensionally reduced flows, besides their own values as the limits of the full-dimensional system under some strong physical constraints, shall also offer useful information for a final reunification of component-wise dimensionally reduced flows to constitute a more complete turbulence theory.

### ACKNOWLEDGMENTS

We thank Dr. H. Zhou for the help with the computer program. PS particularly thanks Prof. W. Zou for supports. The authors acknowledge the support from the computer centre of the Aerospace Institute of Nanchang University where part of the computations were performed.

**DATA AVAILABILITY**

The data that support the findings of this study are available from the corresponding author upon request.

---

helicity effects would be even more promising if united with other techniques such as that fixing the helicity by constructing models for equilibrium and non-equilibrium statistical dynamics [J.-Z. Zhu. "Chiral turbulence: Equilibrium and non-equilibrium ensembles for time-reversible systems," (in Chinese) SCIENTIA SINICA Physica, Mechanica & Astronomica 50, 040002 (2020)].